\documentclass[aps, prx, reprint, superscriptaddress, longbibliography, floatfix]{revtex4-2}
\usepackage{graphicx}
\usepackage{dcolumn}
\usepackage{bm}
\usepackage{physics}
\DeclareMathAlphabet{\pazocal}{OMS}{zplm}{m}{n}
\usepackage{amsmath}
\usepackage{amsfonts}
\usepackage{mathrsfs}
\usepackage{subfigure}
\usepackage{color}
\usepackage{xcolor}
\usepackage[colorlinks=true,linkcolor=blue,anchorcolor=red,citecolor=blue, urlcolor=blue]{hyperref}

\begin{document}
 
\title{Transformer Learning of Chaotic Collective Dynamics in Many-Body Systems}

\author {Ho Jang}
\affiliation{Department of Physics, University of Virginia, Charlottesville, Virginia, 22904, USA}

\author {Gia-Wei Chern}
\affiliation{Department of Physics, University of Virginia, Charlottesville, Virginia, 22904, USA}

\begin{abstract}
Learning reduced descriptions of chaotic many-body dynamics is fundamentally challenging: although microscopic equations are Markovian, collective observables exhibit strong memory and exponential sensitivity to initial conditions and prediction errors. We show that a self-attention–based transformer framework provides an effective approach for modeling such chaotic collective dynamics directly from time-series data. By selectively reweighting long-range temporal correlations, the transformer learns a non-Markovian reduced description that overcomes intrinsic limitations of conventional recurrent architectures. As a concrete demonstration, we study the one-dimensional semiclassical Holstein model, where interaction quenches induce strongly nonlinear and chaotic dynamics of the charge-density-wave order parameter. While pointwise predictions inevitably diverge at long times, the transformer faithfully reproduces the statistical “climate” of the chaos, including temporal correlations and characteristic decay scales. Our results establish self-attention as a powerful mechanism for learning effective reduced dynamics in chaotic many-body systems.
\end{abstract}

\date{\today}
\maketitle

Over the past decade, machine-learning approaches have become powerful tools for learning and forecasting the dynamics of complex quantum and hybrid quantum-classical systems directly from data~\cite{bandyopadhyay2018,yang2020,lin2021,wu2021,rodriguez2021,ullah2021,secor2021,wang2021,lin2022,ullah2022,ullah2022b,rodriguez2022,lin2025,flurin2020,khanahmadi2021,choi2022,koolstra2022,mohseni2022,rodriguez2024,ning2025}. Early efforts focused on feed-forward neural networks and kernel methods to approximate short-time propagators, while recurrent neural networks (RNNs)~\cite{Funahashi93,Mandic01,Cardot11,lipton2015,mienye2024}, particularly long short-term memory (LSTM) architectures~\cite{hochreiter97,sherstinsky20}, were later introduced to capture temporal correlations and history dependence. Much of this progress has been driven by dissipative and open quantum systems, where non-Markovian dynamics arises from coupling to an external environment. In such settings, neural networks effectively learn reduced dynamical maps or implicit memory kernels induced by environmental degrees of freedom. The dynamical regimes considered in most existing studies, however, are relatively simple, typically characterized by relaxation or oscillatory behaviors.

By contrast, closed-system evolution poses a qualitatively distinct and comparatively less explored challenge for data-driven modeling. Although the underlying microscopic equations of motion are deterministic and Markovian---and neural networks have indeed been employed as time-dependent variational states to learn such full Markovian dynamics~\cite{hartmann2019,schmitt2020,reh2021,luo2022,santos2023,medvidovic2023,nys2024,walle2025,zhang2025}---reduced descriptions in terms of collective observables are generically non-Markovian. In a data-driven setting, integrating out microscopic degrees of freedom generates long-ranged temporal correlations and delayed feedback, effectively rendering the collective variables self-interacting through an internal reservoir. The resulting emergent memory reflects information redistribution within the system itself, rather than dissipation or decoherence. 

This challenge is further compounded in chaotic regimes, where exponential sensitivity to initial conditions fundamentally limits long-time trajectory prediction even in the absence of external noise. Previous work has shown that recurrent architectures such as echo state networks (ESNs)~\cite{jaeger04,manjunath13,Chen20,dinh2025} can successfully learn and predict chaotic dynamics when trained on the full system state, where the underlying dynamics remains Markovian and accurate one-step prediction suffices~\cite{Pathak17,Pathak18,Haluszczynski19,Chattopadhyay20,Roy22}. In contrast, for reduced descriptions based on collective observables, chaos manifests through effective non-Markovian dynamics with long memory, rendering pointwise prediction inadequate. Capturing such behavior therefore requires models that can explicitly learn long-range temporal structure and robust statistical regularities, rather than relying solely on local-in-time forecasting.

These challenges have motivated growing interest in transformer architectures with self-attention layers~\cite{Vaswani2017,Wolf2020,Gillioz2020,Radford2023,dosovitskiy2021,Han2023}, originally developed for large-scale language modeling, in which explicit recurrence is replaced by a global attention mechanism. Self-attention enables the model to dynamically reweight correlations across all time steps, providing direct access to long-range temporal dependencies. From a dynamical-systems perspective, attention-based models offer a natural framework for learning reduced descriptions with extended memory, hierarchical temporal organization, and scale-adaptive relevance~\cite{donoso2022,Geneva2022,rodriguez2024}. Recent work has shown that self-attention–based transformers can outperform traditional RNNs in learning dissipative and non-Markovian quantum dynamics from short trajectories, underscoring their promise for modeling chaotic many-body dynamics at the level of collective observables~\cite{zhang2023,Yousif2023,Solera-Rico2024,He2025,Zhai2025}.

\begin{figure*}
    \centering
    \includegraphics[width=0.9\linewidth]{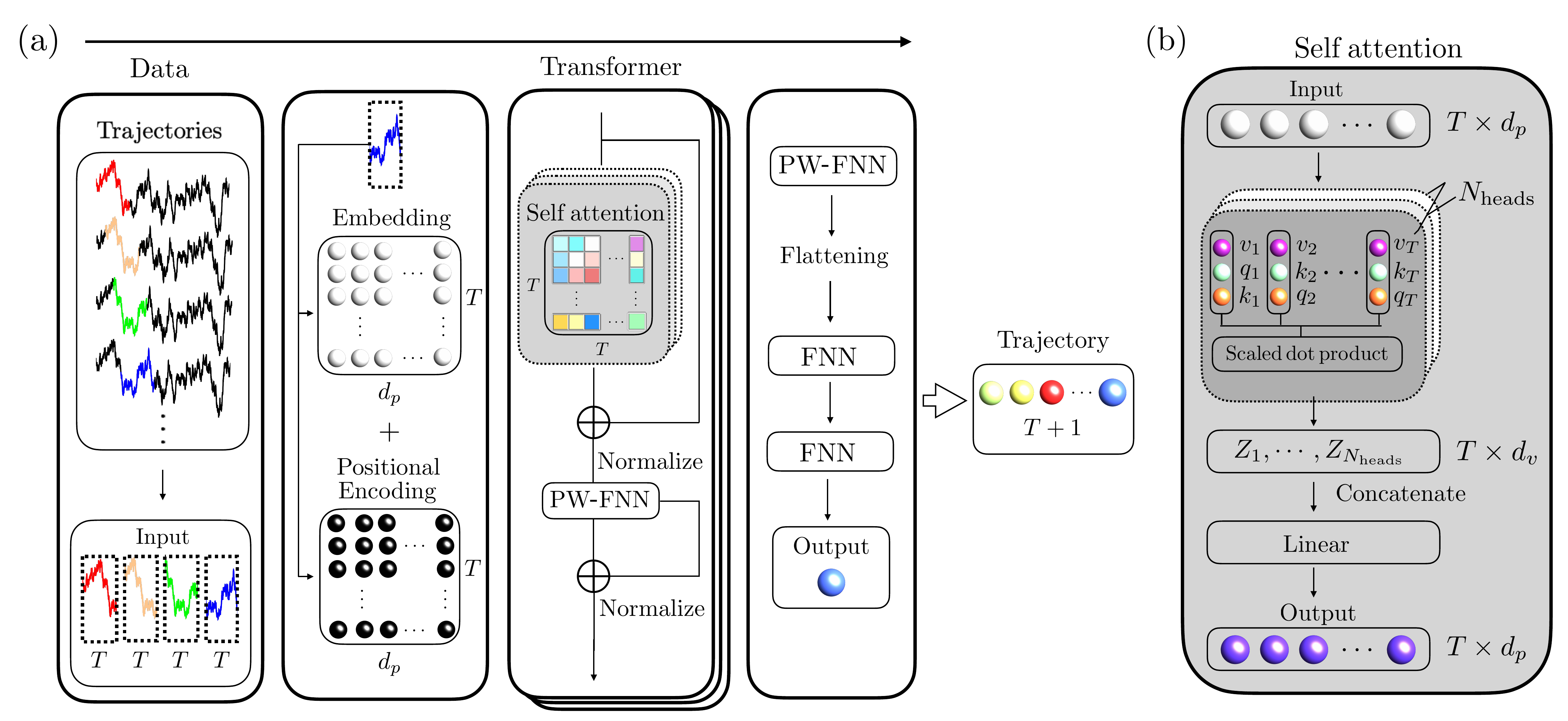}
\caption{(a) A length-$T$ history of the order parameter $\Delta(t)$ is embedded, augmented with positional encoding, and processed by stacked transformer blocks with multi-head self-attention to predict the next value $\Delta(t+\delta t)$. (b)~Multi-head self-attention computes relevance weights between different times via learned query--key interactions, enabling global temporal correlations across the input window.}
    \label{fig:transformer_schematic}
\end{figure*}

We demonstrate that a self-attention–based transformer framework can learn and forecast chaotic collective dynamics in an energy-conserving many-body system. Although the microscopic dynamics are Markovian in full phase space, the reduced dynamics of macroscopic order parameters are intrinsically history dependent. By attending over long temporal contexts, the transformer selectively extracts the correlations most relevant for prediction, enabling accurate reduced descriptions even as individual trajectories diverge at long times due to chaos. As a stringent testbed, we study the one-dimensional semiclassical Holstein model~\cite{holstein59}, where classical lattice dynamics are coupled to quantum electronic evolution. Following an interaction quench, the emergent charge-density-wave order parameter exhibits strongly nonlinear and chaotic behavior. Training the model directly on order-parameter time series, we find that while pointwise trajectory agreement is lost at long times, the transformer faithfully reproduces the statistical “climate” of the dynamics, including temporal correlations and characteristic decay scales. These results highlight self-attention as a physically transparent and effective mechanism for learning reduced descriptions of chaotic many-body dynamics beyond the predictability horizon of individual trajectories.

We first outline a framework for applying transformer models to predict collective dynamics in many-body systems. Without loss of generality, we focus on a single scalar order parameter $\Delta(t)$, defined as the expectation value of an observable $\hat{\mathcal{O}}$, $\Delta(t)=\langle\Psi(t)|\hat{\mathcal{O}}|\Psi(t)\rangle$. In the Heisenberg picture, its evolution follows from $d\Delta/dt=i\langle[\hat{\mathcal{H}},\hat{\mathcal{O}}]\rangle$, where $\hat{\mathcal{H}}$ is the Hamiltonian of the many-body system. Since this commutator cannot, in general, be expressed solely in terms of $\hat{\mathcal{O}}$, the dynamics of $\Delta(t)$ is not closed and is coupled to the dynamics of microscopic degrees of freedom. As a result, the reduced dynamics of the order parameter is effectively non-Markovian, with memory and feedback induced by the many-body environment, as commonly encountered in both interacting quantum systems and hybrid quantum–classical dynamics.

From this perspective, the evolution of $\Delta(t)$ can be viewed as an effective reduced dynamics with an emergent memory kernel, in which past configurations influence future behavior. Even when the full many-body dynamics is Markovian, projection onto a small set of collective variables generically induces history dependence, delayed feedback, and long-lived temporal correlations. At a schematic level, the resulting dynamics may be written in the Mori--Zwanzig form~\cite{Mori1965,Zwanzig1961,Zwanzig2001}
\begin{equation}
	\frac{d\Delta(t)}{dt} = M\,\Delta(t) + \int_0^t K(t-s)\,\Delta(s)\,ds + F(t),
\end{equation}
where $K(t)$ encodes memory effects arising from eliminated degrees of freedom and $F(t)$ represents a fluctuating force determined by unresolved initial conditions. This structure highlights the need for models that can access extended temporal contexts and selectively identify dynamically relevant past information, rather than relying solely on instantaneous or local-in-time predictors.

To address this challenge, we employ a transformer-based model that predicts the next-time-step value of the order parameter from a finite temporal history,
\begin{eqnarray}
	\Delta_{T+1} = \mathcal{F}(\Delta_1, \Delta_2, \cdots, \Delta_T).
\end{eqnarray}
where $\Delta_n=\Delta(t_r+n\delta t)$, $t_r$ is an arbitrary reference time, and $\delta t$ is the time step. Transformer architectures equipped with self-attention naturally meet these requirements. By dynamically reweighting contributions from different time steps, self-attention provides a flexible, data-driven representation of effective memory kernels without imposing a fixed functional form or characteristic timescale. In this sense, the learned attention weights can be viewed as a discrete, adaptive approximation to the memory kernel $K(t)$, encoding how strongly past values of $\Delta$ influence its future evolution.

Fig.~\ref{fig:transformer_schematic} presents a schematic illustration of how this mapping is implemented. The input consists of a windowed sequence of order-parameter values sampled at uniform intervals $\delta t$. Each value is embedded into a higher-dimensional latent space and augmented with positional encoding to retain temporal order. The resulting sequence is processed by a stack of transformer blocks combining self-attention and feed-forward layers, which globally mix information from all past times within the window. This learned temporal coarse-graining enables the model to capture nonlocal correlations and delayed feedback intrinsic to reduced many-body dynamics. The final representation is mapped to a prediction for $\Delta_{T+1}$.

\begin{figure}
    \centering
    \includegraphics[width=0.99\linewidth]{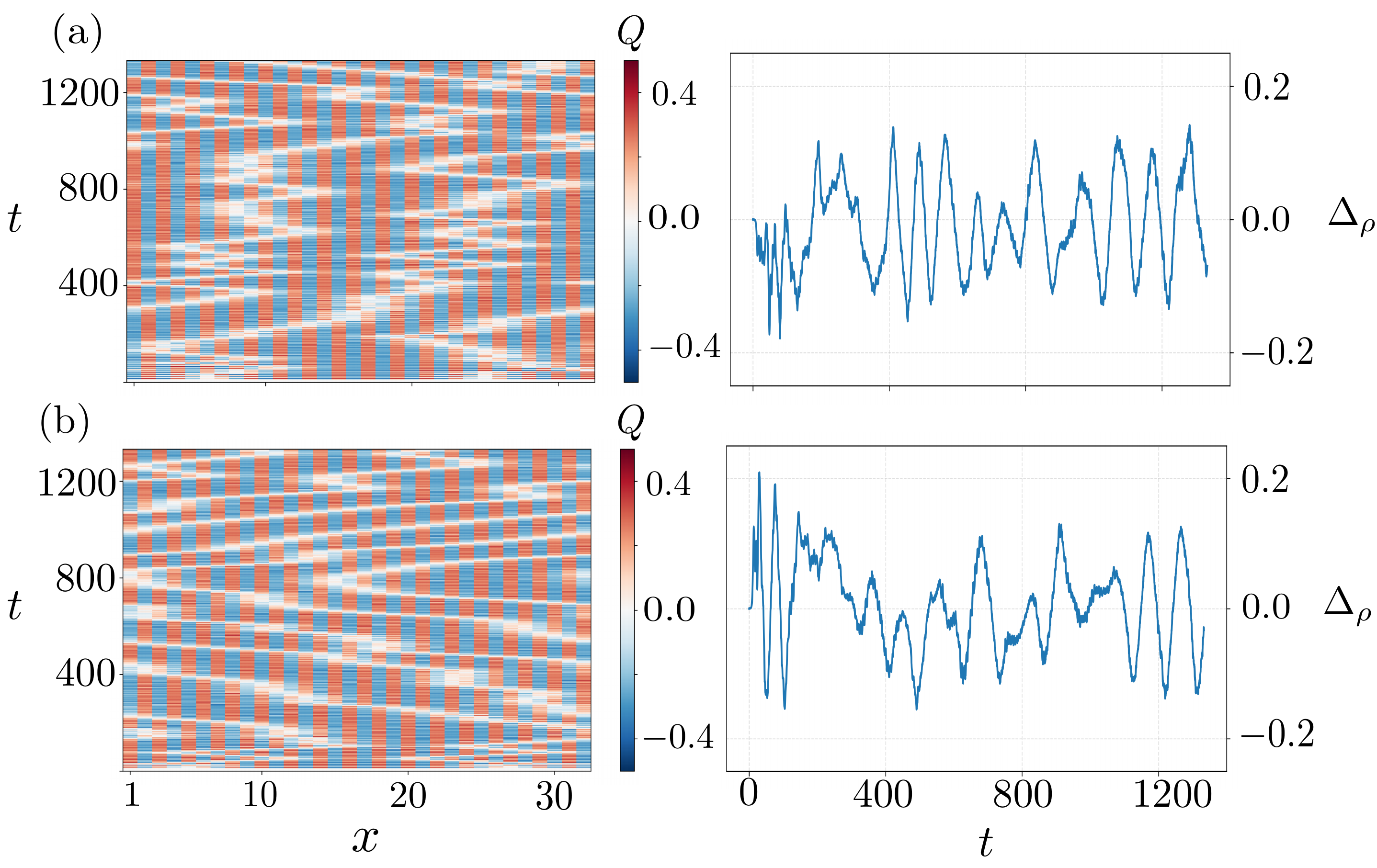}
    \caption{Two representative interaction-quench trajectories of the semiclassical Holstein model. The coordinate $x = i a$ is measured in lattice constant $a$, while time $t$ is measured in unit of time-step, which is set to $\delta t = 0.01 \Omega^{-1}$.  {Left:} spatiotemporal evolution of the lattice displacement field $\{Q_i(t)\}$, showing mobile kink (domain-wall) defects separating CDW domains. {Right:} time evolution of the global CDW order parameter $\Delta_{\rho}(t)$, exhibiting irregular fluctuations driven by the collective kink dynamics.}
    \label{fig:real_trajectory}
\end{figure}

The essential mechanism enabling this behavior is the self-attention operation, illustrated schematically in Fig.~\ref{fig:transformer_schematic}(b). For each time step in the input window, the latent representation is mapped to a set of query, key, and value vectors. Scaled dot products between queries and keys quantify the relevance of different past times to one another, producing attention weights that determine how strongly information is exchanged. In this way, self-attention dynamically constructs an effective, data-driven memory kernel: different regions of the past trajectory can be selectively emphasized or suppressed depending on their relevance for predicting the next-time-step evolution. The use of multiple attention heads allows the model to represent several such correlation structures simultaneously, akin to resolving multiple dynamical modes or relaxation channels within the collective dynamics.


As a proof of principle, we study the post-quench dynamics of charge-density-wave (CDW) order in the one-dimensional Holstein model with spinless fermions~\cite{holstein59}. The Hamiltonian describes nearest-neighbor electron hopping locally coupled to Einstein phonons,
\begin{eqnarray}
	& & \hat{\mathcal{H}} = - t_{\rm nn} \sum_i \left( \hat{c}^\dagger_i \hat{c}_{i+1} + \text{H.c.} \right) - g \sum_i Q_i \hat{n}_i \nonumber \\
	& & \qquad + \sum_i \left( \frac{ P_i^2}{2m} + \frac{m\Omega^2 Q_i^2}{2} \right).
\end{eqnarray}
At half filling, this model supports a robust commensurate CDW phase~\cite{Noack91,Zhang19,Chen19,Esterlis19,Petrovic22} even at the semiclassical level, making it a minimal setting for studying nonequilibrium collective dynamics in a hybrid quantum-classical system.

The dynamical evolution of the system are described by Newton equation of motion for the lattice degrees of freedom coupled to the von~Neumann equation for electrons
\begin{eqnarray}
	m\frac{d^2 Q_i}{dt^2} = g\rho_{ii} - K Q_i, \quad \frac{d\rho}{dt} = -i \bigl[ H(\{Q_i\}), \rho \bigr], \qquad
\end{eqnarray}
where $\rho_{ij}(t)=\langle \hat{c}^\dagger_j \hat{c}_i \rangle$ is the single-particle density matrix and $H_{ij} = -t_{\rm nn}(\delta_{j,i+1}+\delta_{j,i-1}) - g\,\delta_{ij} Q_i(t)$ is the effective single-particle Hamiltonian. Notably, the microscopic equations of motion governing the coupled electron–phonon system are strictly Markovian and conserve a total energy~\cite{yang25,Mitric2024,sankha2024}.

\begin{figure}[t]
    \centering
    \includegraphics[width=0.99\linewidth]{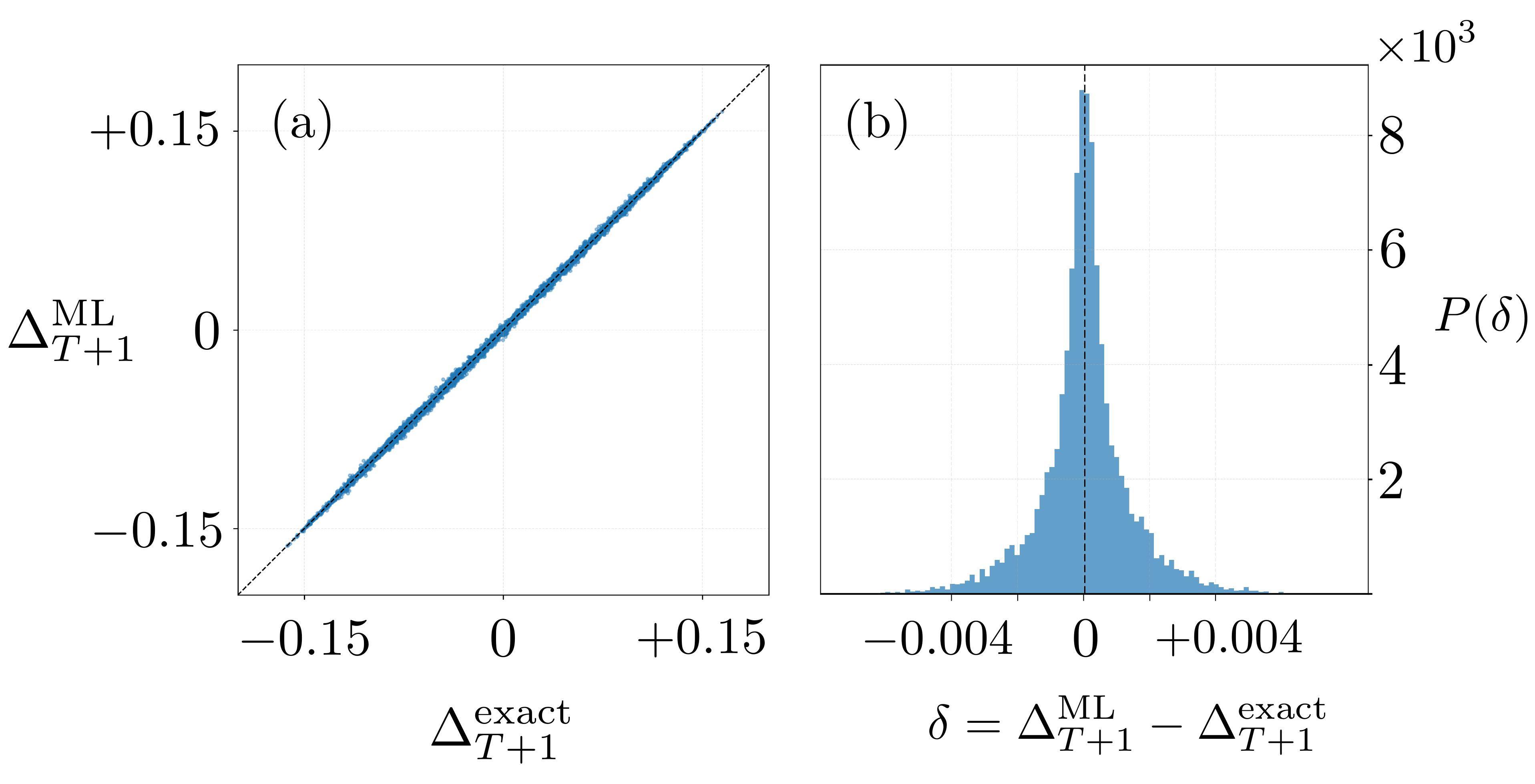}
    \caption{Prediction benchmark of the transformer model.
(a) Scatter plot of the one-step prediction $\Delta_{T+1}^{\rm ML}$ versus the ground-truth value $\Delta_{T+1}^{\rm exact}$, showing excellent agreement along the diagonal. (b) Probability distribution of the prediction error $\delta=\Delta_{T+1}^{\rm ML}-\Delta_{T+1}^{\rm exact}$, which is sharply peaked around zero, indicating high predictive accuracy.}
    \label{fig:validation}
\end{figure}

We focus on the nonequilibrium evolution of the Holstein system following a strong interaction quench, in which initially decoupled electronic and lattice subsystems are suddenly coupled by a finite electron–phonon interaction $g$ at $t\ge 0$. Fig.~\ref{fig:real_trajectory} presents two representative realizations of such $g=0\rightarrow g>0$ quenches. The left panels show the space–time evolution of the lattice displacement field after the quench. Shortly after the interaction is switched on, the system rapidly fragments into multiple charge-density-wave (CDW) domains separated by sharp kink-like domain walls. These kinks propagate through the system, collide, and continuously reorganize, giving rise to a highly irregular spatiotemporal pattern that reflects strong nonlinear interactions and the absence of simple coherent relaxation.

The resulting microscopic complexity is directly reflected in the macroscopic dynamics. The right panels show the corresponding time traces of the staggered charge order parameter, $\Delta_\rho(t)=\sum_i^L n_i(t)(-1)^i$, which exhibit erratic, aperiodic fluctuations characteristic of chaotic collective behavior. As kinks move and interact, local CDW domains transiently grow and shrink, leading to large and irregular variations in $\Delta_\rho(t)$. While the two quenches share the same qualitative phenomenology, their detailed trajectories differ substantially, highlighting the sensitivity of the global order-parameter dynamics to the underlying defect configurations and their nonlinear interactions.

We next apply the transformer model to the time evolution of the CDW order parameter $\Delta_\rho(t)$ generated by the Holstein quench dynamics. Using a finite history window of length $T$ as input, the model is trained to predict the next-time-step value of the order parameter, thereby learning an effective reduced dynamical map for the collective CDW degree of freedom. Fig.~\ref{fig:validation} benchmarks the accuracy of this one-step-ahead prediction. In panel~(a), the predicted values $\Delta^{\rm ML}_{T+1}$ are plotted against the exact results from the Ehrenfest dynamics, showing a tight clustering along the diagonal and indicating excellent agreement over the full range of $\Delta_\rho$. Panel~(b) shows the probability distribution of the prediction error $\delta=\Delta^{\rm ML}_{T+1}-\Delta^{\rm exact}_{T+1}$, which is sharply peaked around zero with a narrow width. Together, these results demonstrate that the transformer accurately captures the short-time evolution of the CDW order parameter despite the strongly nonlinear and chaotic nature of the underlying dynamics, providing a reliable local predictor that forms the basis for longer-time statistical forecasting.

\begin{figure}[t]
    \centering
    \includegraphics[width=0.99\linewidth]{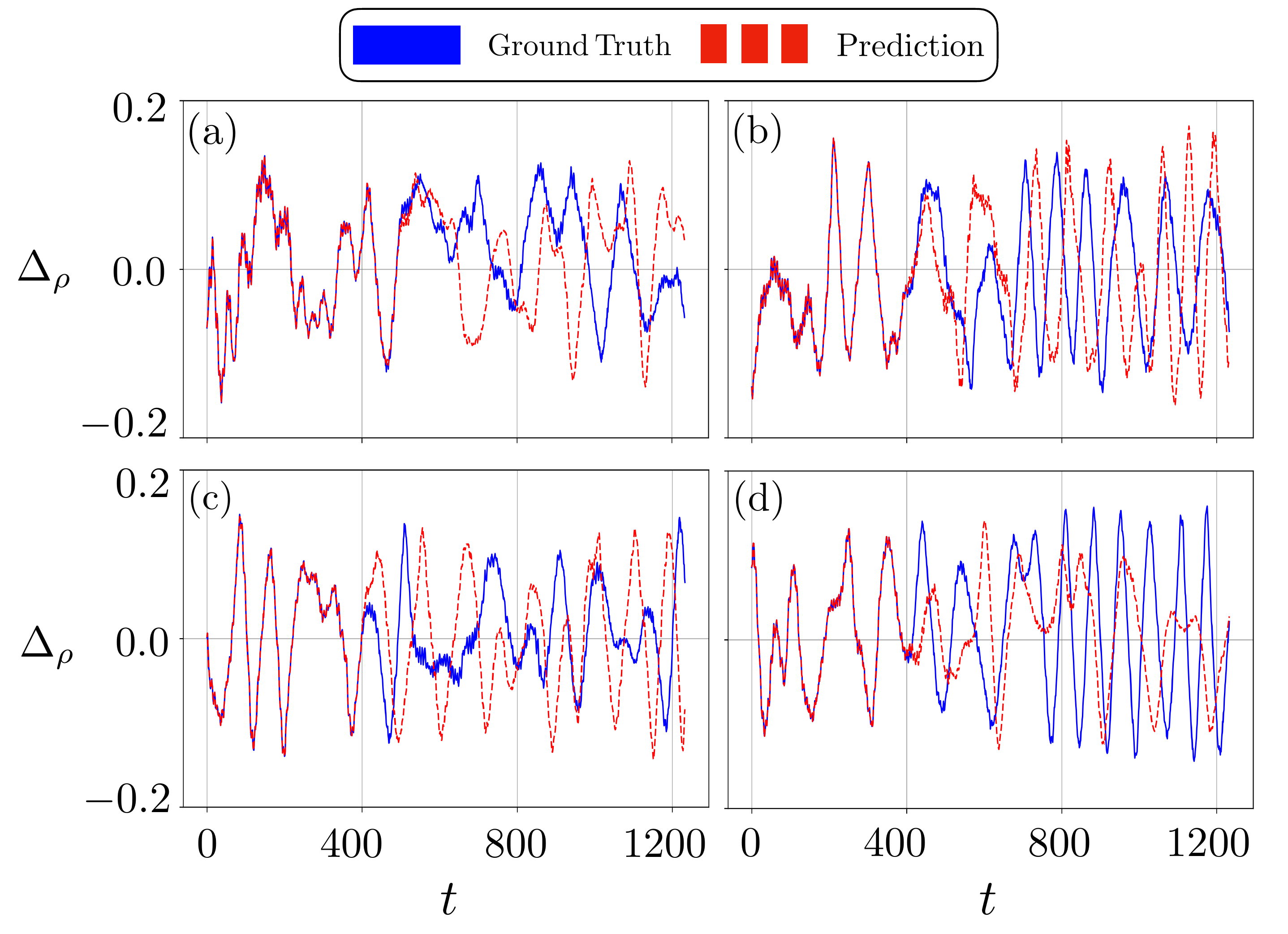}
    \caption{Four representative trajectories of the global CDW order parameter $\Delta_\rho(t)$ from exact simulations (solid blue) and transformer predictions (dashed red). The model shows excellent short-time agreement, with long-time divergence due to error accumulation in the chaotic regime, while correctly capturing the statistical “climate” of the dynamics. }
    \label{fig:gcdw_trajectory}
\end{figure}

Building on this local, one-step-ahead accuracy, we next apply the transformer model to iteratively simulate the full nonequilibrium evolution following a deep interaction quench from $g=0$ to $g>0$. Starting from an initial history window taken from the exact dynamics, the model is recursively advanced in time by feeding its own predictions back as input, thereby generating long trajectories of the global CDW order parameter. Fig.~\ref{fig:gcdw_trajectory} compares four representative trajectories of $\Delta_\rho(t)$ obtained from the transformer and from the exact Ehrenfest simulations. In all cases, the predicted trajectories closely track the exact dynamics at short times, while deviations inevitably accumulate at longer times due to the chaotic nature of the system. Despite this divergence at the level of individual trajectories, the transformer predictions faithfully reproduce the statistical properties of the dynamics, including the characteristic amplitude, fluctuation scale, and temporal variability of $\Delta_\rho(t)$. This demonstrates that, although long-time trajectory prediction is fundamentally limited, the transformer successfully captures the correct dynamical “climate” of the post-quench CDW dynamics.

\begin{figure}[t]
    \centering
    \includegraphics[width=0.95\linewidth]{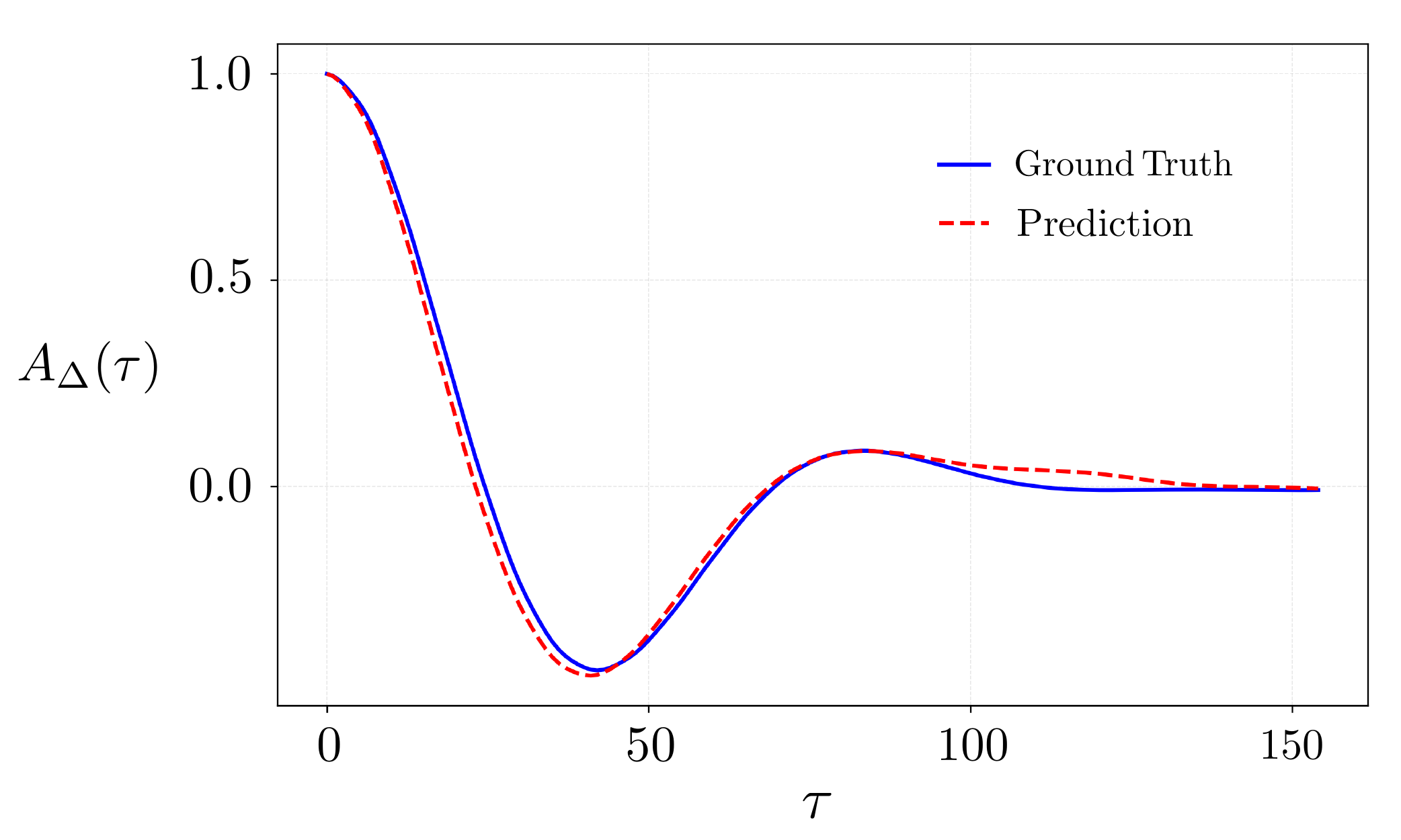}
    \caption{Autocorrelation function of the global CDW order parameter $\Delta_\rho(t)$ computed from exact Ehrenfest dynamics (solid blue) and from transformer-generated trajectories (dashed red).  }
    \label{fig:autocorrelation}
\end{figure}

To quantify more directly whether the transformer captures the statistical properties—or “climate”—of the chaotic CDW dynamics, we compute the autocorrelation function of the global order parameter from long-time trajectories, $A(\tau) = ( \langle \Delta_\rho(t) \Delta_\rho(t+\tau) \rangle - \overline{\Delta}_\rho^2 ) / \sigma_{\Delta_\rho}^2$, where $\overline{\Delta}_\rho$ and $\sigma_{\Delta_\rho}^2$ denote the mean and variance of $\Delta_\rho(t)$, respectively. Figure~\ref{fig:autocorrelation} compares the autocorrelation functions obtained from transformer-generated trajectories and from exact Ehrenfest simulations. Excellent agreement is observed over the entire range of time delays, encompassing the initial rapid decay, the subsequent negative minimum, and the damped oscillatory recovery at longer times. This agreement demonstrates that the transformer accurately reproduces the characteristic correlation timescales and dominant oscillation frequencies underlying the chaotic CDW dynamics.

Importantly, the close match of the autocorrelation functions indicates that the transformer captures not only the overall fluctuation amplitude but also the temporal memory structure encoded in the chaotic dynamics. Since the autocorrelation is insensitive to the precise phase of individual trajectories, this agreement provides a stringent test of the model’s ability to learn the effective reduced dynamics governing the collective CDW degree of freedom. Together with the trajectory-level comparisons, these results establish that, while long-time prediction of individual realizations is fundamentally limited by chaos, the transformer reliably reproduces the correct statistical behavior of the post-quench CDW dynamics. 

To summarize, we have shown that self-attention–based transformer models provide an effective framework for learning reduced descriptions of chaotic collective dynamics in many-body systems directly from time-series data. Using the one-dimensional semiclassical Holstein model as a benchmark, we demonstrated that while long-time prediction of individual trajectories is fundamentally limited by chaos, the transformer accurately reproduces the statistical properties of the dynamics, including temporal correlations and characteristic decay scales of the charge-density-wave order parameter. These results establish self-attention as a powerful mechanism for capturing non-Markovian collective dynamics that emerge from underlying Markovian microscopic equations of motion.

It is instructive to contrast this data-driven approach with time-dependent neural quantum state (NQS) methods based on variational wave-function representations and the time-dependent variational principle~\cite{hartmann2019,schmitt2020,reh2021,luo2022,santos2023,medvidovic2023,nys2024,walle2025,zhang2025}. In the NQS paradigm, a neural network parametrizes the many-body wave function itself, and the dynamics—whether unitary or open—is obtained by approximately solving the underlying equations of motion within a restricted variational manifold. By contrast, the present framework does not attempt to represent the microscopic quantum state. Instead, it learns an effective reduced dynamical map directly from time-series data of collective observables, predicting future evolution from their history. While both approaches can access order-parameter dynamics in closed and open systems, they address complementary objectives: variational NQS methods aim at faithful microscopic state evolution, whereas our framework is tailored to capturing emergent, non-Markovian collective dynamics and their statistical structure without explicit reference to an underlying wave-function description.

The framework introduced here is broadly applicable and admits several natural extensions. It can be generalized to collective variables with more complex internal structure, such as multi-component or spatially resolved order parameters, relevant to systems with competing orders or defect-mediated dynamics. Moreover, because the approach is agnostic to the origin of the training data, it can be readily combined with time series generated by advanced numerical techniques, such as time-dependent density-matrix renormalization group, as well as with data from open many-body systems subject to dissipation and decoherence. Finally, by operating directly on experimentally accessible observables, this framework offers a promising route toward data-driven modeling of nonequilibrium collective dynamics in both quantum and classical many-body experiments.

\bigskip

\textit{Acknowledgments:}~This work was supported by the Owens Family Foundation. G.W.C. thanks Alexei A. Kananenka for insightful discussions on transformer models. The authors acknowledge Research Computing at the University of Virginia for providing computational resources and technical support.

\appendix

\section{Transformer model}

In this Appendix, we provide detailed information on the transformer architecture used in this work, as illustrated schematically in Fig.~\ref{fig:transformer_schematic}. The model is designed to learn the reduced dynamics of the global charge-density-wave (CDW) order parameter following a quantum quench, where the ensuing kink dynamics generates strongly nonlinear and chaotic time dependence. From a single long trajectory $\Delta(t)$ sampled at discrete intervals $\delta t$, we construct a dataset by partitioning the time series into overlapping segments of fixed length $T$. Including the initial point, a trajectory of total length $\mathcal{T} = t_{\mathrm{max}}/\delta t$ yields $\mathcal{T} - T + 1$ such input windows, each of which serves as an independent training example for the transformer.

Each input window consists of a sequence of $T$ scalar order-parameter values, which is mapped into a $T \times d_p$ latent representation through the combination of a value embedding and a positional encoding. The embedding is implemented as a linear feed-forward layer without activation, producing a $d_p$-dimensional vector for each time step. This embedding maps the scalar order parameter into a higher-dimensional feature space where more complex temporal relationships can be expressed. Identical order-parameter values are therefore mapped to identical value embeddings; however, their overall token representations differ once positional encoding is added and contextualized through self-attention. As a result, the model can distinguish identical values occurring at different times or embedded within different dynamical contexts.

Since the embedding layer itself does not encode temporal ordering, explicit time information is introduced through positional encoding. Following Refs.~\cite{Vaswani2017,rodriguez2024,donoso2022}, we employ a trigonometric positional encoding of the form
\begin{equation}
    \mathrm{PE}_{j,k} =
    \begin{cases}
    \sin\,\!\bigl(\omega_k t_j \bigr), & k~\text{even},\\[4pt]
    \cos\,\!\bigl(\omega_k t_j\bigr), & k~\text{odd},
    \end{cases}
    \label{eq:PE-def}
\end{equation}
where $j\in\{1,\ldots,T\}$ labels the position within the input window and $k\in\{1,\ldots,d_p\}$ indexes the feature dimension. The characteristic frequencies are chosen as $\omega_k = 1000^{-2k/d_p}$, ensuring that each embedding dimension probes a distinct temporal scale. A key advantage of this sinusoidal encoding is that it naturally represents relative temporal offsets: for any time shift $\tau$, there exists a linear transformation relating $\mathrm{PE}_{t+\tau}$ and $\mathrm{PE}_{t}$. In our implementation, the positional encoding is constructed from relative times $t_j \in \{0,\delta t,\ldots,(T-1)\delta t\}$, which are identical for every input window and therefore do not encode any absolute time information.

The resulting embedded and positionally encoded sequence $X \in \mathbb{R}^{T \times d_p}$ is then processed by a stack of transformer blocks. The central component of each block is the self-attention mechanism, illustrated schematically in Fig.~\ref{fig:transformer_schematic}(b). For a given input $X$, the output of the $i$th attention head is
\begin{equation}
    Z_i = \mathrm{softmax}\!\left( \frac{Q_i K_i^{\top}}{\sqrt{d_k}} \right) V_i,
    \label{eq:attn-head}
\end{equation}
where the queries, keys, and values are linear projections of the input,
\begin{eqnarray}
	Q_i = X W^{q}_i, \quad  K_i = X W^{k}_i, \quad V_i = X W^{v}_i. 
\end{eqnarray}
Here,
$W^{q}_i, W^{k}_i \in \mathbb{R}^{d_p \times d_k}$ and
$W^{v}_i \in \mathbb{R}^{d_p \times d_v}$ are trainable matrices, while $d_k$ and $d_v$ denote the internal and output dimensions of each attention head. The scaled dot products between queries and keys quantify the relevance of different time steps to one another, and the resulting attention weights determine how information from past times is aggregated.

The outputs of all attention heads are concatenated and projected back to the model dimension,
\[
Z = \mathrm{CONCAT}\{Z_1, Z_2, \ldots, Z_{N_h}\} W^{o},
\]
with $W^{o} \in \mathbb{R}^{N_h d_v \times d_p}$ a trainable projection matrix and $N_h$ the number of attention heads. Each self-attention block is followed by a position-wise feed-forward neural network (PW-FNN) with a single hidden layer with 1536 neurons and a $\tanh$ activation, applied together with residual connections and layer normalization. This combination allows the model to nonlinearly transform the attended representations while maintaining stable gradients during training.

After passing through two such transformer layers, the sequence representation is further processed by a PW-FNN that reduces the feature dimension prior to flattening. The flattened representation is then fed into two fully connected layers with 1024 and 1408 neurons, respectively, both using \texttt{ReLU} activation functions. The final output layer consists of a single neuron, which predicts the next-time-step value of the order parameter,~$\Delta_{T+1}$.

Our implementation builds upon the transformer architecture introduced in Ref.~\cite{rodriguez2024}, which itself closely follows Refs.~\cite{Vaswani2017,donoso2022} but does not employ a separate encoder--decoder structure. Relative to these earlier works, our model incorporates several important modifications tailored to chaotic many-body dynamics. In particular, we find that substantially enlarging the input window $T$ is essential for stable training. For a typical trajectory length $\mathcal{T}=1233$, using small windows such as $T=41$ leads to poor convergence, whereas increasing $T$ to 201 and 401 yields smooth loss decay and dramatically improved performance. A larger window provides a more continuous representation of the underlying dynamics, allowing the model to infer long-range temporal structure rather than relying on sparse local information.

Enlarging the time window necessitates a corresponding increase in the embedding dimension. We find that the embedding size used in previous work ($d_p=64$) is insufficient to represent the richer temporal structure captured by longer windows, leading to collapse onto trivial constant predictions. Systematic tests indicate that $d_p=500$ provides a good balance between expressivity and computational cost, achieving training losses on the order of $10^{-5}$ with reasonable training time.

Finally, we emphasize the importance of using relative positional encoding. After removing early-time transients, the remaining CDW dynamics resides in a chaotic regime without a meaningful absolute time origin. Encoding only relative temporal offsets ensures that the model interprets the input as a sequence governed by internal correlations rather than by external time labels, consistent with the autonomous nature of the underlying dynamics.

In all results presented here, we use $d_k=d_p=500$ and $d_v=d_p/N_h=500$ with a single attention head ($N_h=1$), which is sufficient to demonstrate the capability of the model. The network is implemented in PyTorch and trained using the Adam optimizer with an initial learning rate of $10^{-6}$. We use a batch size of 128 and mean-squared error (MSE) as the loss function. The model is trained for 3200 epochs, achieving validation losses in the range $10^{-5}$–$10^{-6}$. All training is performed on an NVIDIA RTX A6000 GPU.

\section{Non-adiabatic dynamics of semiclassical Holstein model}

In this Appendix, we summarize the Ehrenfest formulation of non-adiabatic semiclassical dynamics employed in our simulations of the Holstein model. This approach provides a controlled and numerically exact description of the coupled electron--phonon dynamics when the lattice degrees of freedom are treated classically while the electronic subsystem remains fully quantum mechanical. In particular, starting from homogeneous charge-density-wave (CDW) initial states—including the limits of a decoupled Fermi gas and Einstein phonons—the resulting equations of motion can be solved efficiently and exactly at the semiclassical level, allowing access to long-time nonlinear and chaotic dynamics.

We consider the standard Holstein Hamiltonian describing spinless fermions locally coupled to dispersionless phonons. Within the Ehrenfest approximation, the total quantum state is assumed to factorize into electronic and lattice components, $\ket{\Gamma(t)} = \ket{\Phi(t)} \otimes \ket{\Psi(t)}$, where $\ket{\Phi(t)}$ and $\ket{\Psi(t)}$ denote the phonon and electron wave functions, respectively. The semiclassical treatment of the lattice further assumes a direct-product structure for the phonon state, $\ket{\Phi(t)} = \prod_i \ket{\phi_i(t)}$, so that each lattice site is described by an independent phonon wave packet. As a consequence, expectation values of phonon operators reduce to classical variables,
\begin{equation}
Q_i(t) \equiv \langle \phi_i(t) | \hat{Q}_i | \phi_i(t) \rangle, \,\,\,
P_i(t) \equiv \langle \phi_i(t) | \hat{P}_i | \phi_i(t) \rangle .
\end{equation}
This product-state assumption neglects entanglement and quantum fluctuations of the lattice degrees of freedom, rendering the approach equivalent to a mean-field treatment of phonons.

The equations of motion for the classical lattice variables are obtained by taking expectation values of the Heisenberg equations using the full product state $\ket{\Gamma(t)}$. Evaluating the commutators yields coupled Hamiltonian equations of motion,
\begin{eqnarray}
\label{eq:newton_eq_app}
\frac{dQ_i}{dt} = \frac{P_i}{m}, \qquad
\frac{dP_i}{dt} = g\, n_i(t) - K Q_i ,
\end{eqnarray}
which correspond to Newton’s equations for harmonic oscillators subject to an additional force proportional to the local electronic density, $n_i(t) = \langle \Gamma(t) | \hat{n}_i | \Gamma(t) \rangle$. Owing to the product structure of the wave function, the density reduces to $n_i(t) = \langle \Psi(t) | \hat{n}_i | \Psi(t) \rangle$, so that the lattice dynamics is driven entirely by the instantaneous electronic configuration. Although the phonon and electron wave functions remain unentangled, the dynamics of the two subsystems are nonlinearly coupled through this self-consistent feedback.

The electronic dynamics is governed by the Schr\"odinger equation,
\begin{equation}
i\hbar \frac{\partial}{\partial t} \ket{\Psi(t)} = \hat{\mathcal{H}}_e[\{Q_i(t)\}] \ket{\Psi(t)},
\end{equation}
where the electronic Hamiltonian depends parametrically on the time-dependent lattice displacements. Since the Holstein Hamiltonian is bilinear in fermionic operators, the many-body electronic state remains a Slater determinant throughout the evolution. Rather than evolving the Slater determinant explicitly, it is numerically advantageous to work with the single-particle density matrix,
\begin{equation}
\label{eq:rho_def_app}
\rho_{ij}(t) = \bra{\Psi(t)} \hat{c}^\dagger_j \hat{c}^{\,}_i \ket{\Psi(t)} .
\end{equation}
The diagonal elements yield the local electron density $n_i(t) = \rho_{ii}(t)$ that enters the lattice equations of motion.

The density matrix evolves under an effective single-particle Hamiltonian defined by
\begin{equation}
\hat{\mathcal{H}}_e
=
\sum_{ij} \hat{c}^\dagger_i \, H_{ij}[\{Q_i(t)\}] \, \hat{c}^{\,}_j ,
\end{equation}
with matrix elements $H_{ij} = -t_{ij} - g \delta_{ij} Q_i(t)$, where $t_{ij} = t_{\rm nn}$ for nearest-neighbor sites and zero otherwise. The evolution of the density matrix is then governed by the von~Neumann equation,
\begin{equation}
\label{eq:von_neumann_app}
i\hbar \frac{d\rho}{dt} = [\rho, H],
\end{equation}
which can be written explicitly as
\begin{eqnarray}
\label{eq:von_neumann_explicit_app}
i\hbar \frac{d\rho_{ij}}{dt}
&=& \sum_k \left( \rho_{ik} t_{kj} - t_{ik} \rho_{kj} \right)
+ g \left( Q_j - Q_i \right) \rho_{ij}. \nonumber \\
\end{eqnarray}
Equations~(\ref{eq:newton_eq_app}) and~(\ref{eq:von_neumann_explicit_app}) together define the Ehrenfest dynamics of the semiclassical Holstein model: a closed set of coupled, nonlinear equations describing classical lattice motion and quantum electronic evolution on equal footing. This approach may be viewed as a single-trajectory limit of the truncated Wigner approximation (TWA)~\cite{Paprotzki_2024,tenBrink_2022}, where quantum fluctuations of the phonon degrees of freedom are neglected. More general multi-trajectory schemes incorporate these fluctuations by sampling initial conditions from the full quantum equilibrium distribution, while retaining the same equations of motion.

The single-trajectory Ehrenfest approach employed here provides an accurate description of the coherent nonequilibrium dynamics of photo-excited Holstein systems, including the emergence of chaotic collective behavior. While quantum fluctuations neglected at this level may influence long-time dynamics and require more advanced many-body techniques for a fully quantitative treatment, the proposed transformer framework is readily applicable to data generated by such higher-fidelity simulations, enabling a unified approach to learning reduced dynamics across different levels of theoretical description.

\subsection{Dimensionless form of the governing equations}

For numerical convenience and clarity, we recast the coupled equations of motion for the electronic and lattice degrees of freedom into a dimensionless form. We choose the nearest-neighbor hopping amplitude $t_{\rm nn}$ as the fundamental energy scale and introduce the associated electronic timescale
\begin{equation}
    \tau_e = {\hbar}/{t_{\rm nn}} ,
\end{equation}
which sets the natural unit of time for the electronic dynamics. Dimensionless time is then defined as
$\tilde{t} = t / \tau_e$, and all energies are measured in units of $t_{\rm nn}$. 

As discussed in the main text, a characteristic scale for the lattice distortion can be identified by balancing the elastic energy $K Q^2$ against the electron–lattice coupling energy $g n Q$, yielding
\begin{eqnarray}
	Q^* = g/K,
\end{eqnarray}
We therefore introduce dimensionless lattice variables $\tilde{Q}_i = Q_i / Q^*$. In terms of the rescaled time $\tilde{t}$ and lattice displacement $\tilde{Q}_i$, the von Neumann equation for the electronic density matrix, takes the form
\begin{equation}
    i \frac{d\rho_{ij}}{d\tilde{t}}
        = \sum_k \bigl( \rho_{ik} \tilde{t}_{kj} - \tilde{t}_{ik} \rho_{kj} \bigr)
          + 4\lambda \bigl(\tilde{Q}_j - \tilde{Q}_i\bigr)\rho_{ij} ,
    \label{eq:A2}
\end{equation}
where $\tilde{t}_{ij} = t_{ij}/t_{\rm nn}$ and $\lambda$ is the dimensionless electron–phonon coupling,
\begin{equation}
    \lambda = {g^2}/{KW}
\end{equation}
with $W = 4 t_{\rm nn}$ denoting the electronic bandwidth. The parameter $\lambda$ thus controls the strength of electron–lattice feedback in the dimensionless equations of motion.

For the lattice sector, the natural timescale is set by the bare phonon frequency $\Omega$, or equivalently
\begin{eqnarray}
	\tau_L = 1/\Omega = \sqrt{m/K}.
\end{eqnarray}
The ratio of lattice and electronic timescales defines the adiabatic parameter
\begin{equation}
    r = \tau_e / \tau_L =  {\hbar \Omega}/{t_{\rm nn}},
\end{equation}
which controls the relative speed of lattice motion compared to electronic dynamics. A characteristic momentum scale for the lattice follows from the relation $dQ/dt = P/m \sim \Omega Q$, giving
\begin{eqnarray}
	P^* = m \Omega Q^*
\end{eqnarray}
Introducing the dimensionless momentum $\tilde{P}_i = P_i / P^*$, the Newton equations of motion for the lattice degrees of freedom can be written in dimensionless form as
\begin{eqnarray}
	\frac{d\tilde{Q}_i}{d \tilde{t}} = r \tilde{P}_i, \qquad 
	\frac{d\tilde{P}}{d \tilde{t}} = -r \tilde{Q}_i + r\left( \rho_{ii} -\frac{1}{2} \right).
\end{eqnarray}

\subsection{Simulation detail}
The deep-quench protocol is implemented using direct numerical simulations of the semiclassical Holstein model. The system is initially prepared in a decoupled state with vanishing electron--phonon coupling ($g_i = 0$), for which the ground state consists of a free electron gas on a one-dimensional lattice and a set of independent harmonic oscillators with $Q_i = P_i = 0$. This initial configuration is spatially homogeneous and does not exhibit charge-density-wave (CDW) order. In the absence of CDW order, inversion symmetry constrains the electronic ground state to be an equal superposition of states at $k_F = \pm \pi/2$, yielding a vanishing net electronic velocity.

At time $t = 0$, the electron--phonon coupling is suddenly switched on to a finite value $g_f > 0$, driving the system far from equilibrium.  Considering any non-vanishing electron-phonon coupling opens the bandgap and make CDW state as ground state, the initial state inject finite energy to the ground state with $g_f$. The ensuing dynamics are governed by the emergence of CDW order induced by the finite post-quench coupling. CDW formation proceeds via a stochastic nucleation process, in which local ordered regions appear at random locations and subsequently grow and merge into extended domains. This behavior closely resembles phase-ordering dynamics following a quench from a disordered state into a symmetry-broken phase and leads to strong sensitivity of the dynamics to microscopic fluctuations in the initial conditions. To control these fluctuations, the initial lattice displacements $Q_i$ are sampled from a normal distribution with zero mean and a standard deviation of~$10^{-4}$. The dynamics are simulated in the dimensionless formulation using parameters $(r,\lambda)=(0.5,0.67)$, and the time evolution is performed using a fourth-order Runge--Kutta scheme with a fixed time step $\delta t = 0.01$.

\bibliography{ref}

\end{document}